\documentclass[aps,prl,preprint,groupedaddress]{revtex4}

\usepackage{graphicx}
\usepackage{amssymb}
\usepackage{epstopdf}
\def\be{\begin{equation}}   \def\ee{\end{equation}}
\def\eq#1{{Eq~(\ref{#1})}}    \def\fig#1{{Fig.\ref{#1}}}

\def\FF{{\cal F}}   \def\eq#1{{Eq.(\ref{#1})}}		\def\PP{{\cal P}}
\def\bd0{{B_d^{(0)}}}     \def\bdmx{{B_d^{(max)}}}  \def\lD{{\lambda_D}}
\def\bs0{{\bar{S_0}}}             \def\sc2mx{{{sc^2}_{|_{max}}}}

\begin{document}

\title{Dynamics of non-equilibrium membrane bud formation}
\author{Pierre Sens}
\affiliation{Institut Charles Sadron - 6 rue Boussingault, 67083 Strasbourg - France}
\affiliation{CNRS/UMR 168 - Institut Curie, section recherche - 11, rue Pierre et Marie Curie
75231 Paris Cedex 05 - France\\Email: Pierre.Sens@curie.fr}

\date{\today}



\begin{abstract}
The dynamical response of a lipid membrane to a {\it local} perturbation of its molecular symmetry is investigated theoretically. A density asymmetry between the two membrane leaflets is predominantly released by in-plane lipid diffusion or membrane curvature, depending upon the spatial extent of the perturbation. It may result in the formation of non-equilibrium structures (buds), for which a dynamical size selection is observed. A preferred size in the $\mu m$ range is predicted, as a signature of the crossover between membrane and solvent dominated dynamical membrane response.
\end{abstract}

\maketitle


The formation of small membrane structures (vesicles or tubules...) is required for most inter and intra cellular transports in biological cells\cite{alberts}. While important progress have been made in the identification of key membrane proteins, recent work have focused on the lipid molecules themselves\cite{lipidmodif1}. Of particular interest is the lipid translocation by specific enzymes (flippases), and the membrane morphological changes they trigger\cite{lipidtransloc2, lipidtransloc1}. From a physical point of view, the formation of a bud from a fluid membrane has mostly been considered as the result of phase separation in mixed membranes\cite{vesicle1}, or as a global shape transition of closed membranes due to geometric frustration\cite{purebud}. The final membrane conformation then corresponds to a global equilibrium, and is not expected to strongly depend upon dynamics\cite{buddyn}. Small invaginations of the plasma membrane of biological cells should however result from localized perturbations, rather than from global changes at the scale of the cell. This paper present a theory for the dynamics of relaxation of one such local perturbation; a local membrane asymmetry, created by a sudden flip of a number of lipid from one leaflet of the bilayer to the other. This situation is of fundamental interest, as it may result in the formation of non-equilibrium membrane structures. The model also seeks to mimic experiments where the local perturbation of a giant vesicle ($\sim 100\mu m$) by adsorption of DNA molecules results in the production of small dynamical vesicles ($\sim\mu$m) with their membrane packed with adsorbed DNA\cite{agelDNA1}.

In the present theoretical analysis, it is shown that the local membrane perturbation introduced by lipid translocation (stretching of the depleted monolayer, and compression of the enriched one, \fig{sketch}), may be released both by opposite monolayer flows, leading to a diffusion of the perturbation (\fig{sketch}a) or by membrane curvature (\fig{sketch}b). While the former mechanism potentially leads to a vanishingly small energy, the faster of the two processes will control the membrane relaxation. It is shown below that large-scale membrane curvature is hindered by bulk flow, while small scale-membrane curvature is slowed by membrane flow. The cross-over between these two regimes defines a critical dynamical lengthscale $\lD\sim 1\mu$m in water and $\sim 100$nm in the more viscous inner cellular environment, at which the perturbation is optimally converted into transient membrane curvature, and fully formed membrane bud are most likely to be observed.

Lipid membranes are self-assembled fluid bilayers. Their equilibrium properties such as  lipid density $\phi_0$ and membrane thickness $2d$ ($\sim 4$nm), result from a balance of the hydrophobic attractions between the lipid tails by steric or electrostatic repulsions\cite{seifert}. Deviation from the optimal density costs an elastic energy ({\em per monolayer}, per unit area) $K_s(\phi-\phi_0)^2/(2\phi_0^2)$, with a large stretching modulus $K_s\sim 25 k_BT/nm^2$\cite{evans} ($k_BT$ is the thermal energy). The membrane bending energy $\kappa C^2/2$ involves the local membrane curvature $C$ and a fairly small bending modulus $\kappa\sim 25 k_BT$. Noting the outer (+) and inner (-) monolayer densities (at the mid-plane) $\psi_\pm$ (with $\phi_\pm=\psi_\pm(1\mp dC)$), the membrane elastic energy is best expressed in terms of the local average dilation $\bar\rho\equiv \frac{\psi_++\psi_-}{2\psi_0}-1$ and the dilation difference between the monolayers $\rho\equiv \frac{\psi_+-\psi_-}{2\psi_0}$\cite{seifert}:
\be
\FF_{elast}=\int dS(K_s(\rho-dC)^2+K_s{\bar\rho}^2+\frac{\kappa}{2}C^2)
\label{enelast}
\ee
Bending a symmetrical bilayer ($\rho=0$) involves a bending modulus $\kappa+2 K_s d^2\sim 200 k_BT$. It is reduced to $\kappa$ in fluid membranes since the monolayer densities can adjust to $\rho=dC$. The (small) ``bending ratio''  of these two quantities $\epsilon\equiv\kappa/(2K_sd^2)$ will prove important later.

Here, we study the membrane response to the sudden establishment of a local concentration asymmetry: some fraction $\rho_0$ of lipid being flipped from the ``down'' to the ``up'' monolayer over an area $S_0$ (\fig{sketch}). For simplicity the membrane bears no global surface tension, which is to say that $\bar\rho=0$ throughout. Lipid flow toward the stretched region of the ``down'' monolayer and away from the compressed region of the ``up'' monolayer, diffuse the perturbation over the membrane (\fig{sketch}a). Concomitantly, the compression-extension  creates a bending torque leading to membrane curvature (\fig{sketch}b). Diffusion of the perturbation is limited by the sliding friction between monolayers, while membrane curvature is limited by membrane and solvent viscous dissipation.

Local balance between elastic and dissipative forces yields local equations for the evolution of the membrane shape. Such equations exist for small membrane deformations\cite{seifert}, but are hardly tractable for large deformations. In consequence, we take a simpler approach where the perturbed membrane (of area $S(t)$) is parameterized by a spherical cap of constant curvature $C(t)$ and constant dilation difference $\rho(t)$, connected to an unperturbed (flat - $C=\rho=0$) membrane, see \fig{sketch}b. This approximation disregards the existence of a connecting neck between the bud and the flat membrane, the stability of which probably influences the scission of the bud by membrane fusion\cite{fusion}. In this work, buds are simply assumed to detach beyond a critical value of the budding parameter (defined in \eq{buddingparam}). In the biological context, it may correspond to a neck small enough for membrane proteins such as dynamin to pinch off the bud\cite{dynamin}. 

Spontaneous lipid flip-flop between the two monolayers is very unlikely over the evolution time, which imposes the conservation of asymmetry $S(t)\rho(t)=S_0\rho_0$. The stretching stress in the perturbed region is released when $C=\rho/d$ (\eq{enelast}). The membrane geometry is described by a {\it budding parameter}, equal to unity for closed buds:
\be
0<B_d\equiv \frac{SC^2}{16\pi}<B_d^{(0)}
\label{buddingparam}
\ee
where $B_d^{(0)}=S_0\rho_0^2/(16\pi d^2)$ corresponds to a situation where the perturbation is entirely converted into curvature, without diffusion along the membrane. As we will see, $B_d$ is always much smaller than $B_d^{(0)}$. The optimal conversion of perturbation into curvature is observed for a particular dynamical lengthscale $\sqrt{S_0}\sim\lD$ (\eq{lD}).

Dynamical equations for the  two independent variables $S$ and $C$ are obtained using a Lagrangian description\cite{goldstein}, where elastic and dissipative ``forces'' are calculated from the variation of the elastic energy $\FF_{elast}[\{\rho,C\}]$, and the energy dissipated per unit time $\PP_{diss}$:
\be
\frac{\partial \FF_{elast}}{\partial\{\rho,C\}}+\frac{\partial \PP_{diss}}{\partial\{d_t\rho,d_t C\}}=0
\label{lagrange}
\ee
Variation with $\rho$ accounts for the migration of lipid molecules under a gradient of chemical potential, and variation with $C$ describes the membrane deformation due to the bending torque.

The three sources of dissipation involve three constitutive parameters. The viscosities of the solvent $\eta$ ($\simeq10^{-3}$N/m$^2$ for water) and of the membrane $\mu_m$ ($\simeq 10^{-9}$N.s/m), couple to gradients of bulk and membrane velocity fields ${\bf v}$ into viscous shear stresses $\sigma_{ij}=\eta\left(\partial_j v_i+\partial_i v_j\right)$\cite{brenner}. An interlayer friction parameter $b_m$ ($\simeq 10^8$N.s/m$^3$), couples to the velocity difference $\delta v$ between membrane the two monolayers in relative motion\cite{evans}. Note that  $\mu_m$ and $b_m$ are related to the membrane (3-dimensional) viscosity $\eta_m$ by the scaling relations $\eta_m\sim \mu_m/ d\sim b_md$, where the membrane viscosity is typically a thousand times the viscosity of water $\eta_m\sim 1$ N/m$^2$\cite{dimova}. The total dissipation is $\PP_{diss}=\PP_{b_m}+\PP_{\mu_m}+\PP_\eta$, with
\be
\PP_{b_m}=\frac{b_m}{2} \int dS\delta v^2\hspace{0.04in}; \hspace{0.04in} \PP_{\eta}=\frac{1}{2\eta}\int dV\sum_{ij}(\sigma_{ij})^2
\label{powerdissip}
\ee
The membrane viscous dissipation $\PP_{\mu_m}$ is obtained by substituting $\mu_m$ to $\eta$ in $\PP_\eta$  and integrating over the membrane surface instead of the volume.

The three contributions to the energy dissipation are calculated as follows (see \cite{fisher} for a related calculation). The bud volume $V$ and neck aperture $L$ satisfy $V=\frac{S^2C}{8\pi}\left(1-\frac{2}{3}B_d\right)$, $\pi L^2=S(1-B_d)$. The curvilinear, radial, and spherical coordinate systems defined \fig{coord} are used to parameterize flows in the curved and flat part of the membrane, and in the bulk, respectively. The differential monolayer velocity $\delta v$ is present in the perturbed part of the membrane. It is axisymmetrical, and relaxes the concentration difference $\rho$ according to the continuity relation\cite{seifert} ${\bf \nabla}_c.\delta{\bf v}=-d_t\rho$, or $\delta v=-4\dot\rho(1-\cos{\psi})/(C\sin{\psi})$ (where the dot represents a time derivative, $\nabla_c$ is the curvilinear gradient along the membrane, and $\psi$ is defined \fig{coord}a. 
The intermonolayer friction part of energy dissipated per unit time is calculated from \eq{powerdissip}:
\be
\PP_{b_m}=\frac{b_m }{4\pi}S^2\dot\rho^2\left(\frac{2}{B_d^2}\left(\log\frac{1}{1-B_d}-B_d\right)\right)
\label{dissipfric}
\ee

We assume incompressible flows in the membrane and the surrounding fluid. Most of the membrane flow comes from bringing membrane area from the flat membrane into the bud. The flow matches the variation of the area $\Delta S=S-\pi L^2=S B_d$, and corresponds to a radial velocity field $v_{membrane}=-d_t(\Delta S)/(2\pi l)$ (where $l$ is defined \fig{coord}b). The main source of solvent flow comes from the variation $\dot V$ of the volume inside the bud, which imposes a flow going through the circular neck of radius $L$. In the spherical coordinate system \fig{coord}c, the velocity field is radial and have the expression\cite{brenner}: $v_{bulk}=\frac{3\dot V}{2\pi r^2}\cos^2{\theta}$.  The bulk and membrane viscous dissipation  are obtained from \eq{powerdissip}
\be
\PP_\eta=\frac{22}{5}\eta\frac{(\dot V)^2}{ \pi L^3}\quad;\quad 
\PP_{\mu_m}=\mu_m\frac{\left(d_t(\Delta S)\right)^2}{\pi L^2}
\label{dissipvolmemb}
\ee
Viscous dissipation in the solvent occurs in both sides of the membrane. $\PP_\eta$ in \eq{dissipvolmemb} is valid for the fluid entering the bud. For the fluid expelled by the bud, the neck size $L$ is approximately replaced by the bud radius $2/C$  beyond hemispherical bud. Extra sources of dissipation, such as the solvent flow inside the bud, and membrane shear flow in the curving membrane, have been checked not to modify the membrane relaxation qualitatively, and are omitted here for clarity.
The evolution of $S$ and $C$ is entirely determined by Eqs(\ref{enelast}, \ref{lagrange}, \ref{dissipfric}, \ref{dissipvolmemb}). Notwithstanding the simple assumptions for the membrane shape,  these equations (Eqs.(\ref{fulleq1},\ref{fulleq2})) are fairly complex and have to be dealt with numerically. In the limit of small deformations $B_d\ll1$, all equations may be linearized, and membrane viscous dissipation is negligible.  The membrane asymmetry relaxes via diffusion along the membrane according to the expression $S\dot\rho=-D(\rho-dC)$ with the diffusion coefficient $D$ and  characteristic relaxation time $t_0$ ($\sim 0.1$ms for $S_0\sim \mu$m$^2$):
\be
D\equiv 4\pi\frac{K_s}{b_m}\sim10^{-8}\ {\rm m^2/s}\quad;\quad t_0\equiv\frac{S_0}{D}
\label{diffusion}
\ee
The influence of the perturbation lengthscale $S_0$ is emphasized by using dimensionless variables for area, curvature and time:
\be
s\equiv \frac{S}{S_0}\quad;\quad c\equiv\frac{dC}{\rho_0}\quad;\quad \tau\equiv\frac{t}{t_0}
\label{var}
\ee
The new variables are bounded by: $1<s<\infty$ and $0<c<1$, and the budding parameter is $B_d=\bd0 sc^2$ (see \eq{buddingparam}). Balancing the curvature forces lead to an equation for $C$. The interplay between membrane and solvent dissipation defines a characteristic lengthscale $\lD\sim bd^2/\eta$\cite{seifert, evans, sens_electroform},  to which the size of the perturbation should be compared. 
\be
\lD\equiv\frac{20\sqrt{\pi}}{11}\frac{bd^2}{\eta}\quad;\quad\bs0\equiv\frac{S_0}{\lD^2}
\label{lD}
\ee
This lengthscale  $\lD$ is of order $1\mu$m in water, but is much shorter ($\sim 100$nm) in biological condition, since the cytosol can be very viscous.

Expressed in the dimensionless variables, the set of equations reads\cite{validation}:
\be
\left(\frac{2}{B_d^2}\left(\log\frac{1}{1-B_d}-B_d\right)\right)\dot s=(1-sc)
\label{fulleq1}
\ee

\begin{eqnarray}
\left[\sqrt\bs0\frac{(1-2 B_d)^2}{(1-B_d)^{3/2}}\left(s^{5/2}\dot c+2\frac{(1-B_d)}{(1-2B_d)}s^{3/2} c\dot s\right)\right]\cr
+\left(\frac{\mu_m}{b_md^2}\right)\frac{B_d}{(1-B_d)}\left(s^2\dot c+sc\dot s\right)=(1-(1+\epsilon)sc)
\label{fulleq2}
\end{eqnarray} 
 The driving forces for diffusion and membrane deformation (RHS of the above equations) both show the competition between increases of $s$ and $c$. 
 
 At short time ($s\sim 1$ and $c\ll 1$), the evolution is given by $\dot s\sim1$ and $\dot c\sim 1/\sqrt{\bs0}$. For $\bs0<1$, membrane curvature moves little solvent and occurs faster than diffusion. On the other hand, membrane deformation is slow for $\bs0>1$. For larger deformation however, the membrane viscosity comes into play. It is characterized by a single parameter $\alpha_m\equiv\mu_m/(b_md^2)$ (typically slightly larger than unity\cite{dimova}), and becomes important for small scale perturbations ($B_d\gtrsim\sqrt{\bs0}/\alpha_m$). 
 
Eqs.({\ref{fulleq1},\ref{fulleq2}) show that membrane deformation will evolve non-monotonously. Indeed, the driving force for diffusion ($(1-sc)$ - \eq{fulleq1}) vanishes for complete release of the stretching stress ($sc=1$, or $C=\rho/d$). On the other hand, the driving force for membrane curvature ($(1-(1+\epsilon)sc)$ - \eq{fulleq2}) vanishes for smaller deformation because of membrane bending rigidity ($\epsilon\equiv\kappa/(2K_sd^2)$). As a consequence, the budding parameter rises from zero to a maximum value $\bdmx$ over a time $t_{growth}$, then it decays slowly to zero over a larger time $t_{decay}$. The value of the maximum is a measure of the amount of initial perturbation converted into curvature. It will be discussed extensively in the remaining of this letter, for it determines whether the blister may turn into a well-formed bud before the perturbation diffuses out. For small deformations, the rise occurs over the linear evolution time $t_{growth}\sim t_0$, and the decay, mostly driven by bending rigidity, is of order $t_{decay}\sim t_0/\epsilon$. These times are however much larger in the non-linear regime. To give a feel for the numbers; for $\bs0=1$, the maximum deformation is an hemispherical buds  ($\bdmx=0.5$) for an initial perturbation $\bd0=3$, while nearly completed buds ($\bdmx=0.9$) require $\bd0\gtrsim6$ (see \fig{results}).  Fixing $\lambda_D=1\mu$m, hemispherical buds have evolution times $t_{growth}=0.7$ms and $t_{decay}=26$ms, and nearly completed buds $t_{growth}=7$ms and $t_{decay}=300$ms.

Many results can be extracted from Eqs.(\ref{fulleq1},\ref{fulleq2}), which determines (within the approximations) the full dynamics of the membrane deformation. Here we concentrate on the following question. How much should we perturb the membrane in order to see buds? \fig{results} shows the value of the initial density of flips, and of the initial perturbation $\bd0$ needed to obtain a given maximum of the budding parameter $B_d^*$, as a function of $\bs0$. One may practically picture  $B_d^*$ as some critical value at which the neck connecting the bud to the mother membrane becomes unstable. More generally, it illustrates the non-trivial competition between membrane and solvent dynamics for large membrane deformation. 

\fig{results} shows that membrane viscosity strongly reduces the likeliness  of large membrane deformations ($B_d^*>0.5$) for small initial extensions $\bs0<1$. More extended perturbations on the other hand, are dominated by solvent dissipation. Interestingly in this case, the diffusion of the perturbation (increase of $S$) promotes bud closure (increase of $C$), to minimize the variation of the bud volume ($\dot V=0$ when $S\dot C( 2B_d-1)=2C\dot S(1-B_d)$). In consequence, a bud is likely to close if it reaches the hemispherical shape, and the results in \fig{results} overlap for $\bs0\gg1$. Furthermore, the density needed to reach a given membrane deformation $B_d^*$ surprisingly tends to a constant value, so the total perturbation mass $\rho_0S_0$ linearly increase, as the extension $S_0$ increases.

The interplay between membrane and solvent dissipations produces a minimum of $\bd0$ for a given $S_0$ of order $\lD^2$,  purely dictated by dynamics. Mature buds have best chances to be produced at this particular lengthscale, as the maximum membrane deformation $B_d^*$ is closest to the optimal one $\bd0$. Interestingly, the DNA-induced budding of giant vesicles\cite{agelDNA1} results in endosomes of size close to this optimal dynamical size in water ($\sim\mu$m). It is hazardous to make more quantitative comparison with these experiments, since a number of other factors such as the kinetics of DNA adsorption certainly influences the formation of buds, but it is one prediction of the present work that the bud size can be altered by changing the solvent viscosity.

According to \fig{results}, well-formed buds with $B_d=0.9$ require flipping only $\rho_0\sim 3.5\%$ lipids over an area $S_0\sim 1\mu m^2$ (for $\lD=1\mu m$), while it necessitates $\rho_0=35\%$ for $S_0\sim (200 nm)^2$. This amounts to flipping about $10^5$ molecules in both cases. The flips need not being perfectly synchronized, as successive events may act cooperatively if they are separated by a time lag shorter than the perturbation decay time $t_{decay}$. As already discussed, $t_{decay}\sim t_0/\epsilon\sim S_0/(D\epsilon)$ in the limit of small membrane deformation. Note that the evolution becomes much slower in the large deformation regime, where dissipation is very large. 

In conclusion,  non-linear dynamical equations for the formation of bud-like invaginations in fluid membranes have been derived, and applied to the relaxation of a localized perturbation of the up/down symmetry of lipid bilayers. Non-equilibrium buds may be produced provided that relaxation by membrane curvature occurs faster than the diffusion of the perturbation along the membrane. The membrane response critically depends upon the extension of the initial perturbation, which determines whether the membrane deformation is limited by membrane or solvent  dynamics. Bud formation is most likely at the crossover between the two regimes.

\begin{acknowledgments}
I thank Albert Johner for many enjoyable discussions.
\end{acknowledgments}



\newpage

\begin{thebibliography}{99}

\vskip-2truecm

\bibitem{alberts} B. Alberts, D. Bray, J. Lewis, M. Raff, K. Roberts, J. Watson {\it Molecular
Biology of the Cell.}, Garland, New York (1994)

\bibitem{lipidmodif1} R. Golsteyn {\em Trends Cell Biol.}, {\bf 10} (2000)

\bibitem{lipidtransloc2} E. Farge, D.M. Ojcius, A. Subtil, A. Dautry-Varsat{\em Am. J. Physiol. Cell Physiol.} {\bf 276} C725 (1999) \qquad P.F. Devaux {\em Biochimie} {\bf 82} 497 (2000)

\bibitem{lipidtransloc1} E.M. Bevers, P. Comfurius, D.W. Dekkers and R.F.  Zwaal {\em Biochim. Biophys. Acta} {\bf 1439} 317 (1999)

\bibitem{vesicle1} D.D. Lasic, U.R. Joannic, B.C. Keller, P.M. Frederik, L. Auvray {\em Adv. Colloid Interface Sci} {\bf 89} 337 (2001)\qquad F. Julicher and R. Lipowsky {\em Phys. Rev. Let.} {\bf 70} 2954 (1993)

\bibitem{purebud} H.-G. D\"obereiner, E. Evans, M. Kraus, U. Seifert and M. Wortis {\em Phys. Rev. E} {\bf 55} 4458 (1997)

\bibitem{buddyn}   P.B. Sunil Kumar and M. Rao {\em Phys. Rev. Let.} {\bf 80} 2489 (1998)\qquad P.B. Sunil Kumar,G. Gompper and R. Lipowsky {\em Phys. Rev. Let} {\bf 86} 3911 (2001)

\bibitem{agelDNA1} M.I. Angelova and N.H. Tsoneva {\em Eur. Biophys. J.} {\bf 28} 142 (1999)

\bibitem{seifert} U. Seifert and S.A. Langer {\em Europhys. Lett} {\bf 23}, 71
(1993)

\bibitem{evans} E. Evans and A. Yeung {\em Chem. Phys. Lipids} {\bf 73}, 39
(1994)

\bibitem{sens_electroform} P. Sens and H. Isambert {\em Phys. Rev. Let.} {\bf 88}, 128102 (2002)
               

\bibitem{fusion} R. Jahn and H. Grubm\"uller {\em Curr. Opin. Cell Biol.} {\bf 14} 488 (2002)

\bibitem{dynamin} M.A. McNiven, H. Cao, K.R. Pitts and Y. Yoon {\em TIBS}, {\bf 25} 115 (2000)

\bibitem{goldstein} H. Goldstein {\em Classical Mechanics}, Addison-Wesley 2nd Ed. (1980)

\bibitem{brenner} J. Happel and H. Brenner {\em Low Reynolds Number Hydrodynamics} (1991) Kluwer, Dordrecht

\bibitem{dimova} R. Dimova, B. Pouligny and C. Dietrich {\em Biophys. J.} {\bf 79} 340 (2000)

\bibitem{fisher} T.M. Fisher {\em Phys. Rev. A} {\bf 50} 4156 (1994)

\bibitem{validation} In the small deformation regime, this set of equations reproduce the results of the rigorous linear equations of ref.\cite{seifert}, which validates to some extent the present parameterization of the membrane shape. 

\end{thebibliography}

\newpage

\begin{figure}
 \vspace{8cm}
\centerline {\includegraphics[width=13cm]{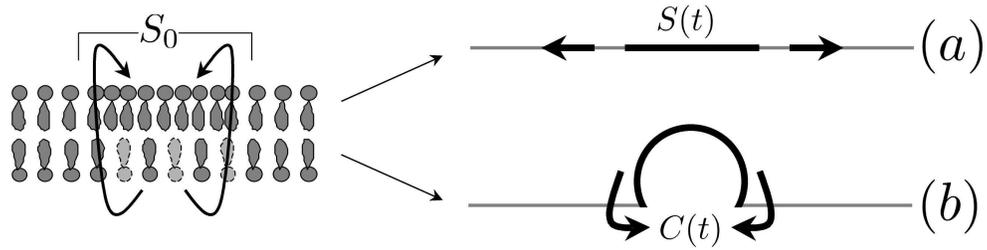}}
\vspace{1cm}
\caption{\label{sketch}Relaxation of a membrane asymmetry (a) in-plane diffusion and (b) membrane curvature.}
\end{figure}

\begin{figure}
\vspace{10cm}
\centerline {\includegraphics[width=13cm]{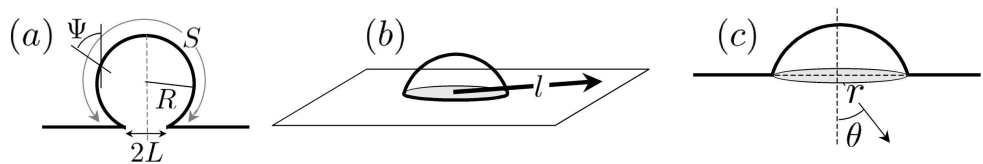}}
\vspace{1cm}
\caption{\label{coord}Parameterization for the membrane shape surface $S$, curvature $C=2/R$ and radius of the neck $L$, and the various coordinate system used to derive the dissipation}
\end{figure}

\begin{figure}
\vspace{7cm}
\centerline {\includegraphics[width=13cm]{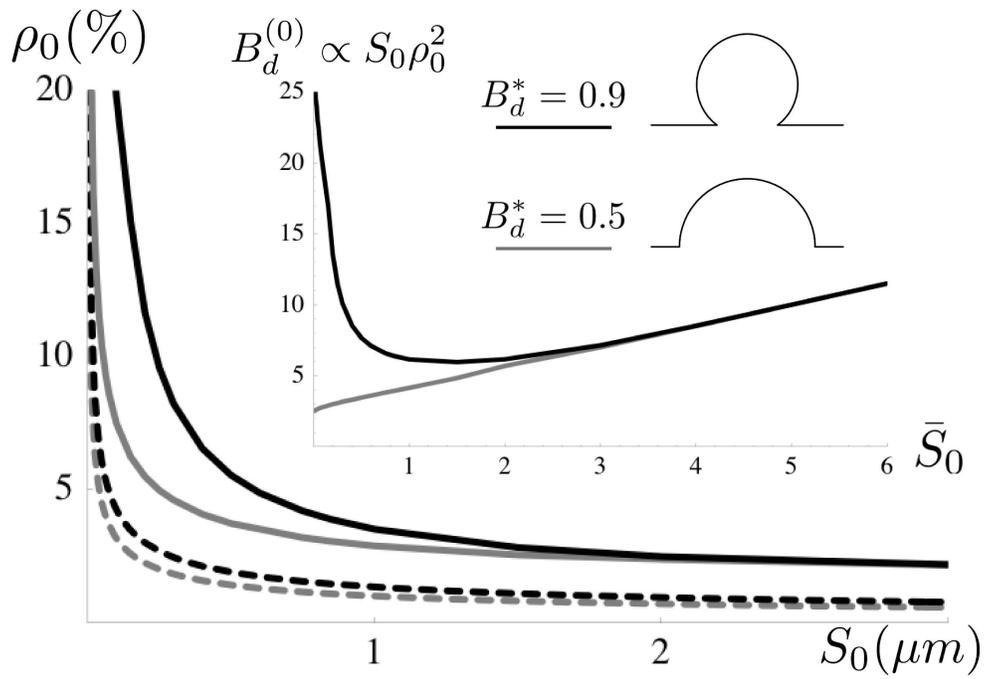}}
\vspace{1cm}
\caption{\label{results}Density of flips $\rho_0$ and strength of the initial perturbation $\bd0$ needed to reach hemispherical and nearly-completed buds, as a function of the initial extension $S_0$ ($\epsilon=0.2$ - $\frac{\mu_m}{b_md^2}=3$ - $\lD=1\mu$m). Dashed lines show the densities $\rho_0$ corresponding to $\bd0=0.5$ (grey) and $0.9$ (black).}
\end{figure}

\end{document}